\documentclass[
reprint,
superscriptaddress,
showpacs,
nofootinbib,
amsmath,amssymb,
aps,
pra,
floatfix,longbibliography
]{revtex4-2}

\usepackage[usenames, dvipsnames]{xcolor}
\usepackage{amsmath, amsthm, amssymb, dsfont}
\usepackage{amsfonts}
\usepackage{times}
\usepackage{txfonts}
\usepackage{enumerate}
\usepackage[english]{babel}
\usepackage{graphicx}
\usepackage[caption=false]{subfig}
\usepackage{bbm}
\usepackage[normalem]{ulem}
\usepackage{color}
\usepackage{hyperref}
\hypersetup{
 pdfnewwindow=true,
 colorlinks=true,
 linkcolor=black,
 citecolor=blue!70!black,
 filecolor=blue!70!black,
 urlcolor=blue!70!black
}

\newcommand{\bra}[1]{\langle #1|}
\newcommand{\ket}[1]{|#1\rangle}

\newcommand{\ketbra}[2]{| #1 \rangle \langle #2 |}

\newcommand{\id}{\mathbbm{1}}
\newcommand{\bea}{\begin{eqnarray}}
\newcommand{\eea}{\end{eqnarray}}

\newcommand{\alex}[1]{{\leavevmode\color{blue}#1}}

\DeclareMathOperator{\Tr}{Tr}

\newtheoremstyle{customStyle1}  
{0pt}       
{0pt}       
{\normalfont}   
{\parindent}        
{\em}  
{. --}   	 
{.5em}       
{\thmname{#1}\thmnumber{ #2}\thmnote{ (#3)}}  

\theoremstyle{customStyle1}

\theoremstyle{plain} 

\newtheorem*{thm*}{\protect\theoremname}
\newtheorem*{prop*}{\protect\propositionname}
\newtheorem*{lem*}{\protect\lemmaname}
\newtheorem*{cor*}{\protect\corollaryname}

\ifx\proof\undefined
\newenvironment{proof}[1][\protect\proofname]{\par
	\normalfont\topsep6\p@\@plus6\p@\relax
	\trivlist
	\itemindent\parindent
	\item[\hskip\labelsep\scshape #1]\ignorespaces
}{%
	\endtrivlist\@endpefalse
}
\providecommand{\proofname}{Proof}
\fi

\let\alex\relax

\begin{document}

\title{Experimental progress on quantum coherence: detection, quantification, and manipulation}

\author{Kang-Da Wu}
\affiliation{CAS Key Laboratory of Quantum Information, University of Science and Technology of China, \\ Hefei 230026, People's Republic of China}
\affiliation{CAS Center For Excellence in Quantum Information and Quantum Physics, University of Science and Technology of China, Hefei, 230026, People's Republic of China}

\author{Alexander Streltsov}
\affiliation{Centre for Quantum Optical Technologies, Centre of New Technologies,
University of Warsaw, Banacha 2c, 02-097 Warsaw, Poland}

\author{Bartosz Regula}
\affiliation{School of Physical and Mathematical Sciences, Nanyang Technological University, 637371 Singapore}

\author{Guo-Yong Xiang}

\email{gyxiang@ustc.edu.cn}
\affiliation{CAS Key Laboratory of Quantum Information, University of Science and Technology of China, \\ Hefei 230026, People's Republic of China}
\affiliation{CAS Center For Excellence in Quantum Information and Quantum Physics, University of Science and Technology of China, Hefei, 230026, People's Republic of China}

\author{Chuan-Feng Li}

\affiliation{CAS Key Laboratory of Quantum Information, University of Science and Technology of China, \\ Hefei 230026, People's Republic of China}
\affiliation{CAS Center For Excellence in Quantum Information and Quantum Physics, University of Science and Technology of China, Hefei, 230026, People's Republic of China}

\author{Guang-Can Guo}

\affiliation{CAS Key Laboratory of Quantum Information, University of Science and Technology of China, \\ Hefei 230026, People's Republic of China}
\affiliation{CAS Center For Excellence in Quantum Information and Quantum Physics, University of Science and Technology of China, Hefei, 230026, People's Republic of China}

\begin{abstract}
Quantum coherence is a fundamental property of quantum systems, separating quantum from classical physics. Recently, there has been significant interest in the characterization of quantum coherence as a resource, investigating how coherence can be extracted and used for quantum technological applications. In this work we review the progress of this research, focusing in particular on recent experimental efforts. After a brief review of the underlying theory we discuss the main platforms for realizing the experiments: linear optics, nuclear magnetic resonance, and superconducting systems. We then consider experimental detection and quantification of coherence, experimental state conversion and coherence distillation, and experiments investigating the dynamics of quantum coherence. We also review experiments exploring the connections between coherence and uncertainty relations, path information, and coherence of operations and measurements. Experimental efforts on multipartite and multilevel coherence are also discussed.
\end{abstract}

\maketitle

   \tableofcontents

\section{Introduction}
The phenomenon of quantum coherence --- or superposition --- represents a fundamental difference between the quantum and classical worlds. It forms the foundation of quantum effects such as multi-particle interference and quantum entanglement, and it continues to play a key role in the application of quantum physics and quantum information science. Although the origins of the study of quantum coherence lie in the efforts to understand properties of optical fields, the recent years have seen the development of general frameworks of superposition between orthogonal quantum states~\cite{aberg2006quantifying,BaumgratzPhysRevLett.113.140401,StreltsovRevModPhys.89.041003,HU20181}, focused in particular on finite-dimensional quantum systems.

The need to characterize the role that coherence plays in driving applications in quantum information science and quantum technologies motivates a precise understanding of several aspects of this phenomenon: its \emph{detection}, that is, the ability to certify when a state or device is truly classical and when it possesses quantum coherence; its \emph{quantification}, that is, measuring and comparing the strength of coherence in different settings; and its \emph{manipulation}, that is, our capability to effectively transform and use coherence in applications.
The resource-theoretic framework~\cite{chitambar2018quantum} provides rigorous tools to describe quantum coherence in analogy
with methods used to study quantum entanglement 
and other nonclassical resources.

The formulation of a resource theory of quantum coherence goes back to~\cite{aberg2006quantifying,BaumgratzPhysRevLett.113.140401}, an approach which attracted significant attention, both on the theoretical and experimental side~\cite{StreltsovRevModPhys.89.041003}. Coherence quantification, and in particular the development of coherence quantifiers which have an operational meaning in experiments, has seen steady active progress within the resource-theoretic framework~\cite{PhysRevLett.116.150502,YuanPhysRevA.92.022124,MarvianPhysRevA.94.052324,ZhuhuangPRA2017,chin_2017-1,chin_2017,ProofRobustness,PhysRevLett.115.020403,Biswas20170170,EstimatingTong2018,WinterPhysRevLett.116.120404,TakagiPhysRevLett.122.140402,uola_2020-1,ducuara_2020}. Other important questions concern state transformation, i.e., the possibility to transform a quantum state into another one by using quantum operations which do not create coherence~\cite{ChitambarPhysRevLett.117.030401,ChitambarPhysRevA.94.052336,YadinPhysRevX.6.041028,WinterPhysRevLett.116.120404,Matera2058-9565-1-1-01LT01,SciRepCoherenceTransQubit,PhysRevLett.119.140402,PhysRevA.91.052120,Du:2015:CMO:2871378.2871381,Regula1711.10512,zhao_2018,PhysRevLett.121.070404,zhao_2019,lami_2019-1,DeterministicCoherenceDistillation}. Coherence theory in multipartite system has also been investigated, allowing to provide powerful links between the theories of entanglement and coherence~\cite{PhysRevLett.115.020403,ChitambarPhysRevLett.116.070402,StreltsovPhysRevX.7.011024,regula_2018-2,ConvertingPRL,Nathan,Vijayan1804.06554}. Various applications rely on the presence of quantum coherence, including quantum sensing and metrology~\cite{MarvianPhysRevA.94.052324,chaoPRAevolution,giorda2017coherence} and quantum computation~\cite{HilleryPhysRevA.93.012111,Matera2058-9565-1-1-01LT01}. Quantum coherence was also shown to play an important role in quantum thermodynamics~\cite{ThermalBrand,gour2015resource,goold2016role,PhysRevXTimeTranslation,lostaglio2015description,lostaglio2017thermodynamic,faist2015gibbs,LimitationsEvolutionPRL,narasimhachar2015low,korzekwa2016extraction,kammerlander2016coherence,solina1,solina2}, and for transport phenomena relevant in biological systems~\cite{huelga2013vibrations,Vatasescu1,Vatasescu1E,BirgittaTransport1,o2014non,engel2007evidence,collini2010coherently,rebentrost2009role,rebentrost2009environment}. 

In this work, we aim to overview the recent developments in the practical study of quantum coherence as a resource, and in particular the experimental implementations of coherence detection, quantification, and manipulation protocols.

We begin in Section~\ref{sec:theory} with an introduction to the theoretical underpinnings of the resource-theoretic approach to quantum coherence, before proceeding to review the experimental progress on various aspects of characterizing and applying coherence in Section~\ref{sec:Experimental}.
 Conclusions are provided in Section~\ref{sec:Conclusions}.

\section{Theoretical background: Resource theory of quantum coherence}\label{sec:theory}
\subsection{Incoherent states}
A quantum resource theory is based on two main ingredients, the free states and the free operations \cite{doi:10.1142/S0217979213450197,chitambar2018quantum,StreltsovRevModPhys.89.041003}, both arising from physical restrictions on the set of quantum operations implementable within the given setting. 
In the case of coherence theory, the free states --- states which do not possess any resource and are cheap to prepare --- are the incoherent states, i.e., quantum states which are diagonal in a fixed reference basis. Thus, the resource theory of quantum coherence explicitly depends on the chosen basis, which we will take to be some complete, orthogonal, and normalized reference basis $\{\ket{i}\}$. This choice is typically based on the actual physical setting, such as the polarization states of photons ($\{\ket{H},\,\ket{V}\}$) or the energy eigenstates of Hamiltonian of a quantum system. In a given $d$-dimensional Hilbert space $ \mathcal{H}$, we will use $\mathcal{I}$ to denote the set of all incoherent quantum states, that is, states which can be written in diagonal form as
\begin{equation}\label{Eq:incoherentstate}
\delta=\sum_{i=0}^{d-1} p_i\ketbra{i}{i},
\end{equation}
where $\sum_{i=0}^{d-1}p_i=1$, $p_i\geq0$ for all $i$.

In multipartite quantum systems, the reference basis is typically chosen as the tensor product of the local reference bases of each subsystem~\cite{theoryFrozen1,PhysRevLett.115.020403,StreltsovRevModPhys.89.041003}. Consider $n$ quantum systems, each with corresponding Hilbert space $\mathcal{H}_{k}$ and local reference basis $\{\ket{i_k}\}$. A pure incoherent state is then a tensor product of local incoherent states: $\ket{i_1} \otimes \ket{j_2} \otimes \cdots \otimes \ket{l_n}$. Any general multipartite incoherent state can be written as a convex combination of such pure incoherent states.

We stress here that the theory of coherence discussed in this work --- although conceptually related --- is distinct from  the notion of ``coherence'' as found in the study of optical nonclassicality based on the Glauber--Sudarshan coherent states~\cite{glauber_1963,sudarshan_1963}, often encountered in optical setups. Efforts have been made to extend the resource theory of quantum coherence to the setting of continuous-variable quantum information~\cite{zhang_2016,regula_2020-1,XuPhysRevA.93.032111,AlbarelliPhysRevA.98.052350} and even to explicitly connect the study of optical nonclassicality to the resource theory of quantum coherence~\cite{tan_2017}, but the two theories are fundamentally different and typically require specialized approaches. Here, we limit ourselves to the discussion of quantum coherence in finite-dimensional spaces, which is how it is most commonly investigated.

\subsection{Incoherent operations \label{sec:Operations}}

In order to study the dynamical behavior of coherence in quantum systems, it is necessary to understand how it evolves under the action of suitable transformations --- the free operations. Ideally, one would like to understand such channels as the operations naturally determined by the physical consideration of the given resource. For instance, in the theory of entanglement, the basic setting of spatially separated laboratories singles out the class of local operations and classical communication (LOCC) as the fundamental set of free operations~\cite{HorodeckiRevModPhys.81.865} owing to the fact that such operations can be performed without using any entanglement between the distant parties. However, an issue arises in the study of quantum coherence, as the structure of this resource theory fails to identify a single, physically motivated class of operations which should serve as the primary class of free transformations. To address this problem, two main approaches have been employed.

On the one hand, one can try to define the free operations through axiomatic considerations. This is done by understanding the natural constraints that a class of free operations should satisfy and studying the sets of all channels obeying such properties.
The basic assumption of this type is that free operations should not generate any coherence; the largest class of such coherence non-generating channels are the \textit{maximally incoherent operations} (MIO)~\cite{aberg2006quantifying}, defined as all channels such that any $\Lambda(\sigma)$ is incoherent for incoherent $\sigma$. This constraint can be made stronger by imposing that the operations should not generate any coherence in any measurement outcome, that is, that $\Lambda$ can be written as $\Lambda(\rho) = \sum_{n} K_n \rho K_n^\dagger$ where each of the Kraus operators $K_n$ satisfies
\begin{equation}
    \frac{K_n\delta K_n^\dag}{\mathrm{Tr}(K_n\delta K_n^\dag)}\in \mathcal{I}\quad\forall\, \delta\in\mathcal{I},\, n. \label{eq:IO}
\end{equation}
Such channels are known as the \textit{incoherent operations} (IO)~\cite{BaumgratzPhysRevLett.113.140401}. Yet another approach is to impose that the operations should not be able to \textit{detect} the coherence of the input state and use this coherence in the transformation. This leads to the class of \textit{dephasing-covariant incoherent operations} (DIO)~\cite{ChitambarPhysRevLett.117.030401,MarvianPhysRevA.94.052324}, which are channels such that $\Lambda \circ \Delta = \Delta \circ \Lambda$, where $\Delta[\rho] = \sum_i \ket{i}\!\bra{i}\rho\ket{i}\!\bra{i}$ denotes complete dephasing in the incoherent basis. If this is imposed at the level of each Kraus operator $K_n$, this then leads to \textit{strictly incoherent operations} (SIO)~\cite{WinterPhysRevLett.116.120404,YadinPhysRevX.6.041028}. Various other constraints can be imposed similarly~\cite{MarvianPhysRevA.94.052324,ChitambarPhysRevA.94.052336,vicente_2017,regula_2020-4}, altogether leading to a diverse range of classes of transformations (see Ref.~\cite{StreltsovRevModPhys.89.041003} for an overview). Although such choices lead to a convenient mathematical description, a common issue with operations defined axiomatically is that neither their physical meaning nor their practical implementability are clear.

On the other hand, the problem can be approached from the other direction, and one can instead ask: what kind of operations can we implement in practice without having to expend any coherence? Clearly, such an approach could provide a natural bridge to experimental implementations of free operations, making it very appealing from both a theoretical and an applied perspective. This motivated the definition of the set of physically incoherent operations (PIO)~\cite{ChitambarPhysRevLett.117.030401}, which can be realized in practice with only incoherent unitary operations and incoherent measurements.
The class SIO also admits similar implementation schemes --- requiring only incoherent ancillae, although more general measurements and unitary transformations~\cite{YadinPhysRevX.6.041028} --- making them another type of experimentally-friendly free operations.  However, such maps were found to be rather limited in their operational power: for instance, both PIO and SIO are incapable of distilling any coherence from general quantum states~\cite{zhao_2019,lami_2019-1,lami_2019}. This suggests that it is difficult to reconcile operational usefulness of free operations with their physical implementability.

Because of these issues, understanding the precise power and limitations of each type of transformations has attracted significant attention. Fortunately, a number of useful equivalences emerged: in many relevant operational tasks, the different sets of transformations were found to have equal capabilities. For instance: the sets DIO, IO, and SIO have exactly the same power in deterministic pure-to-pure quantum state manipulation~\cite{PhysRevA.91.052120,ChitambarPhysRevA.94.052336,regula_2020-4,ZhuhuangPRA2017}; MIO and DIO are both equally powerful in one-shot coherence distillation~\cite{Regula1711.10512}, while IO and SIO have the same power in coherence dilution~\cite{zhao_2018}; DIO and SIO are equally capable in probabilistic pure-state distillation protocols~\cite{PhysRevLett.121.070404}; in the single-qubit case, SIO and IO perform all probabilistic transformations equally well~\cite{wu2020quantum}. Most importantly, the class SIO can even match the performance of the largest class of free operations MIO: this happens in coherence distillation from pure states~\cite{Regula1711.10512}, in the task of environment-assisted coherence distillation of any state~\cite{Regula1807.04705}, and in the deterministic manipulation of all single-qubit systems~\cite{ChitambarPhysRevA.94.052336,PhysRevLett.119.140402}. 
Similar equivalences can be found in the bipartite setting of distributed coherence~\cite{StreltsovPhysRevX.7.011024,yamasaki_2019}, which we will discuss in more detail in Section~\ref{sec:Experimental}.

The results described above lead to the important conclusion that, in many settings, it suffices to consider easily implementable classes of transformations such as SIO, since even larger classes of operations cannot perform any better. This alleviates the issues associated with the different possible choices of transformations to consider, allowing experimental implementations of coherence manipulation schemes to more easily exploit the full transformative potential of free operations. Indeed, most of the tasks discussed in this work will be examples of such operational equivalences, meaning that optimal manipulation schemes can be devised using structurally simpler operations.

\subsection{Quantifying coherence \label{sec:QuantifyingCoherence}}
We will now review methods for coherence quantification, focusing in particular on quantifiers which have an experimental implementation. Any measure of coherence $C$ is nonnegative, zero on all incoherent states, and nonincreasing under any free operation $\Lambda$~\cite{BaumgratzPhysRevLett.113.140401,StreltsovRevModPhys.89.041003}: 
\begin{equation}
    C(\Lambda[\rho]) \leq C(\rho). \label{eq:Monotonicity}
\end{equation}
For simplicity, we consider the class of incoherent operations IO in our discussion here, as this is the most commonly studied choice of free transformations in this context. Another desirable feature of coherence measures is to be non-increasing \textit{on average} under incoherent operations~\cite{BaumgratzPhysRevLett.113.140401}:
\begin{equation}
    \sum_i q_i C(\sigma_i) \leq C(\rho), \label{eq:StrongMonotonicity}
\end{equation}
with probabilities $q_i$ and states $\sigma_i = K_i \rho K_i^\dagger / q_i$ obtained from $\rho$ via an incoherent operation. Condition~(\ref{eq:StrongMonotonicity}) is also called strong monotonicity, and means that coherence does increase under incoherent measurements with postselection on the measurement outcomes. Additionally, many coherence measures discussed in the literature are convex, i.e., $C(\sum_i p_i \rho_i) \leq \sum_i p_i C(\rho_i)$. Note that any convex coherence measure which fulfills Eq.~(\ref{eq:StrongMonotonicity}) gives an upper bound on the optimal state conversion probability~\cite{vidal_2000,wu2020quantum}:
\begin{equation}
    P(\rho \rightarrow \sigma) \leq \min \left\{\frac{C(\rho)}{C(\sigma)},1 \right\},
\end{equation}
where $P(\rho \rightarrow \sigma)$ is the maximal probability for converting $\rho$ into $\sigma$ via incoherent operations.

For any distance measure between quantum states $\mathcal{D}(\rho,\sigma)$, a coherence quantifier can be defined as~\cite{BaumgratzPhysRevLett.113.140401}
\begin{equation}\label{eq:distance}
    C_{\mathcal{D}}(\rho)=\min_{\delta\in\mathcal{I}}\mathcal{D}(\rho,\delta), 
\end{equation} 
corresponding to the minimal distance of $\rho$ to the set of incoherent states $\mathcal{I}$. Condition~(\ref{eq:Monotonicity}) is automatically satisfied for any contractive distance $\mathcal{D}$, i.e. any distance which fulfills $\mathcal{D}(\Lambda[\rho],\Lambda[\sigma]) \leq \mathcal{D} (\rho,\sigma)$ for any quantum operation~$\Lambda$. An important example is the relative entropy of coherence~\cite{BaumgratzPhysRevLett.113.140401}
\begin{equation}
    C_r(\rho) = \min_{\delta \in \mathcal{I}} S(\rho||\delta) \label{eq:Cr}
\end{equation}
with the quantum relative entropy 
\begin{equation}
    S(\rho||\delta)=\Tr (\rho\log_2\rho)-\Tr (\rho\log_2\delta).
\end{equation}
The optimal incoherent state in Eq.~(\ref{eq:Cr}) is found to be $\delta = \Delta[\rho]$, which allows to express the relative entropy of coherence as~\cite{BaumgratzPhysRevLett.113.140401}
\begin{equation}\label{eq:relent}
    C_r(\rho)=S[\Delta(\rho)]-S(\rho),
\end{equation}
where $S(\rho) = -\Tr[\rho \log_2 \rho]$ is the von Neumann entropy. The relative entropy of coherence fulfills all requirements for coherence measures mentioned above~\cite{BaumgratzPhysRevLett.113.140401}. It has an operational meaning in the asymptotic setting, corresponding to the rate of maximally coherent states $\ket{+} = (\ket{0} + \ket{1})/\sqrt{2}$ obtainable via incoherent operations from asymptotically many copies of a state $\rho$~\cite{WinterPhysRevLett.116.120404}.

Other examples for distance-based coherence measures are $\ell_1$-norm of coherence $C_{\ell_1}$~\cite{BaumgratzPhysRevLett.113.140401} and geometric coherence $C_g$~\cite{PhysRevLett.115.020403}:
\begin{align}
    C_{\ell_1}(\rho) &= \min_{\delta\in\mathcal{I}}\|\rho-\delta\|_{\ell_1}, \\
    C_g(\rho) &= 1 - \max_{\delta \in \mathcal{I}} F(\rho,\sigma)
\end{align}
with $\ell_1$-norm $||M||_{\ell_1} = \sum_{i,j}|M_{ij}|$ and fidelity $F(\rho,\sigma) = \left(\mathrm{Tr}\left[\sqrt{\sqrt{\rho}\sigma \sqrt{\rho}}\right]\right)^2$. Both quantities fulfill all requirements for coherence measures discussed above~\cite{BaumgratzPhysRevLett.113.140401,PhysRevLett.115.020403}. Moreover, $C_{\ell_1}$ can be expressed as $C_{\ell_1}(\rho) = \sum_{i\neq j}|\rho_{ij}|$. 

Another class of coherence measures can be obtained via the convex roof construction~\cite{aberg2006quantifying,YuanPhysRevA.92.022124,StreltsovRevModPhys.89.041003}
\begin{equation}
    C(\rho) = \inf \sum_i p_i C(\ket{\psi_i}),
\end{equation}
where $C(\ket{\psi})$ is a suitably defined coherence measure for pure states, and the infimum is taken over all pure state decompositions of $\rho$. An important example is the coherence of formation~\cite{aberg2006quantifying,WinterPhysRevLett.116.120404,YuanPhysRevA.92.022124}
\begin{equation}
    C_f(\rho) = \min \sum_i p_i S(\Delta[\ket{\psi_i}\!\bra{\psi_i}]).
\end{equation}
Coherence of formation has an operational meaning in asymptotic coherence theory, where it corresponds to the rate of maximally coherent states required to create asymptotically many copies of a state $\rho$~\cite{WinterPhysRevLett.116.120404}. Geometric coherence $C_g$ can also be regarded as a convex roof coherence measure, i.e. it fulfills $C_g(\rho) = \min \sum_i p_i C_g (\ket{\psi_i})$~\cite{StreltsovRevModPhys.89.041003}.

Finally, we will also consider the robustness of coherence~\cite{PhysRevLett.116.150502,ProofRobustness}
\begin{equation}\label{eq:robustness_primal}
    C_{R}(\rho)=\min_{\tau}\left\{s \geq 0\,:\,\frac{\rho+s\tau}{1+s} \in\mathcal{I}\right\},
\end{equation}
which can be interpreted as the minimal weight for mixing a quantum state $\rho$ with another state $\tau$ to make it incoherent. Robustness of coherence fulfills all requirements for coherence measures and has an operational meaning in channel discrimination tasks~\cite{PhysRevLett.116.150502,ProofRobustness}.

This completes our review on the theoretical aspects of the resource theory of coherence. In the next section we will review the experimental progress on coherence theory, discussing also methods for experimental estimation of the coherence measures introduced above.

\section{Experimental progress on the resource theory of quantum coherence}
\label{sec:Experimental}

Implementing relevant protocols in quantum resource theories requires experimental platforms satisfying several crucial conditions, such as the ability to prepare high-fidelity states, implementation of sophisticated (often non-unital) quantum channels, and reliable quantum state reconstruction technologies. The three main experimental platforms for realizing quantum resource control and state conversion are linear optics, nuclear magnetic resonance (NMR), and superconducting systems. We will now briefly introduce these three platforms.

{Photonics plays an important role in describing the foundations of quantum mechanics~\cite{MultiphotonPan} and more recently has led to advances in understanding new computational possibilities~\cite{carolan2015universal,zhong2020quantum}. Encoding qubits in the polarization degree of photons has been particularly appealing for the ability to implement arbitrary linear operations via combinations of wave plates~\cite{LOComputingRMP}. Regarding path encoding, the same linear operations can be realized via beamsplitters and phase shifters. Because any linear optical circuit can be described by a unitary operator and a specific array of elementary mode operations is sufficient to implement a unitary operator on different optical modes~\cite{arbitraryunitaryZelinger}, it is theoretically possible to construct a single device with sufficient versatility to efficiently implement any feasible operation up to the given number of optical modes. }

The basic principle of NMR is to use the spin-$\frac{1}{2}$ nucleus as the qubit, where the spin up and spin down states are taken as the systems of interest, i.e., the states $\ket{0}$ and $\ket{1}$ of the qubit, respectively. The single qubit unitary gate can be realized by electromagnetic wave. The energy of the electromagnetic wave and the energy difference between different states of nuclear spin are the same, which is used to control the transformation between the states $\ket{0}$ and $\ket{1}$. A molecule with multiple nuclear spins is regarded as a quantum information processor. Different kinds of nuclei can be used as different qubits since different nuclei have different transition frequencies, so they can use different frequencies of electromagnetic waves to achieve independent manipulation of single bit, that is, addressing. NMR samples often contain a large ensemble of similar molecules, and each molecule is a quantum information processor. 
Since the NMR signal of a single molecule is very weak, all molecules in the sample are often manipulated simultaneously, and the processing results of all molecules are read out, which is called ensemble quantum computation~\cite{RMPNMR}.

{Superconducting circuits are macroscopic systems based on electric oscillators, composed of a large number of atoms in the shape of wire and metal plate. Despite their macroscopicity, they still exhibit general quantum properties, such as quantized energy levels, superposition and entanglement~\cite{clarke2008superconducting,plantenberg2007demonstration}. The working principle of superconducting qubits is based on two phenomena: superconductivity, that is, the frictionless flow of electric fluid through metal at low temperature, and the Josephson effect, which allows nonlinear transformation without introducing dissipation or dephasing.}

Linear optical, NMR, and superconducting systems are the most well-developed technologies for implementing quantum gates with few qubits. They all have distinct properties and excel in different settings. For instance, long distance transmission with high fidelity is an advantage for linear optical systems over the other two systems. However, the interaction between two photons is weak, which means that building multi-qubit gates is more difficult compared to NMR and superconducting systems. For example, the two-photon controlled-NOT gate has a success probability of $1/9$~\cite{linearopticCNOT1,LOComputingRMP}, while for NMR and superconducting qubit, they can be deterministic~\cite{RMPNMR,plantenberg2007demonstration}. As such, none of the three platforms appears to be universally more suitable than the others, and indeed all three have found different applications in the study of quantum information, including quantum coherence.

In the following, we review recent advances in experimental detection and quantification of quantum coherence. 

\subsection{Experimental detection and quantification of coherence}

A straightforward method to detect and quantify coherence in experiments is to
perform state tomography. The density matrix obtained in this way can then be used to evaluate the corresponding coherence measure using analytical and numerical methods, as discussed in Sec.~\ref{sec:theory}. Following this method, the essential task for detection and estimation of coherence is the reconstruction of the quantum state. Quantum state tomography has become a standard technology for inferring the state of a quantum system through appropriate measurements and estimation~\cite{paris2004QSE,PRAmeasurementqubit,CompressedQST,salvail2013full}. To reconstruct a quantum state, measurements are performed on identically prepared copies of a quantum system. The measurement statistic collected in this way then serves as a basis to estimate the density matrix elements using an appropriate algorithm. Examples for such algorithms are maximum-likelihood estimation~\cite{MLEPRL,MLEPRL1,MLEPRL2}, Bayesian mean estimation~\cite{BEPRA}, and least-squares inversion~\cite{LSPRA}. 

The above method is reliable and is based on well-established techniques for quantum state tomography. However, it is not the most efficient, e.g.\ when it comes to the precision of coherence estimation, taking into consideration the number of copies of the quantum state required to collect the statistics. One reason for this is that quantum state tomography reveals the complete information about a quantum state, while the number of measurements grows exponentially with the number of qubits. This makes it difficult to estimate properties such as coherence from the experimental quantum state. On the other hand, only partial information is needed to estimate quantum coherence of a state, a fact which is used by several experimental methods, leading to a higher precision for coherence estimation.

A general approach that dispenses with the need for full-state tomography is to employ \textit{coherence witnesses}. A witness is simply an observable $W$ which satisfies $\Tr(\sigma W) \geq 0$ for all incoherent $\sigma \in \mathcal{I}$. The expectation value of any such witness can then be used to verify that a given system has coherence:  measuring $\Tr(\rho W) < 0$ for some state $\rho$ certifies that $\rho \notin \mathcal{I}$. Originally conceived to detect the presence of entanglement in quantum systems~\cite{terhal_2002}, witnesses were later adapted to provide quantitative bounds for entanglement measures~\cite{brandao_2005,eisert_2007}, and this approach was also introduced to coherence theory~\cite{EstimatingTong2018}. Crucially, for some monotones, an optimization over a carefully chosen set of witnesses can yield the precise value of the measure. A representative example of this is the robustness of coherence $C_R$ defined in Eq.~(\ref{eq:robustness_primal}). It is possible to explicitly express $C_R$ in terms of coherence witnesses~\cite{PhysRevLett.116.150502,ProofRobustness}:
\begin{equation}\begin{aligned}\label{eq:robustness_dual}
  C_{R}(\rho) = \max \left\{ -\Tr (\rho W) \;:\; W \in \mathcal{W},\, \|W\|_\infty \leq 1 \right\}
\end{aligned}\end{equation}
where $\mathcal{W}$ denotes the set of all coherence witnesses and $\|\cdot\|_\infty$ is the operator norm (largest singular value). Therefore, there always exists a witness $W$ such that $C_{R}(\rho) = -\Tr(\rho W)$, allowing for the value of the robustness to be efficiently measured in experiment without the need for state tomography. Importantly, \textit{any} coherence witness can provide a bound for the robustness simply by rescaling it by $\|W\|_\infty$. More accurate bounds can then be obtained by measuring several witnesses, not necessarily tailored to a given state $\rho$, and then optimizing the combined experimental data to yield the best quantitative bound which is compatible with the obtained measurement results (see Ref.~\cite{ProofRobustness}). Such an approach can also be adapted to provide bounds for other coherence measures~\cite{eisert_2007,EstimatingTong2018,Untrusted2,WitnessUntrustedPRL2019}.

\begin{figure}
	\centering
	\includegraphics[scale=0.6]{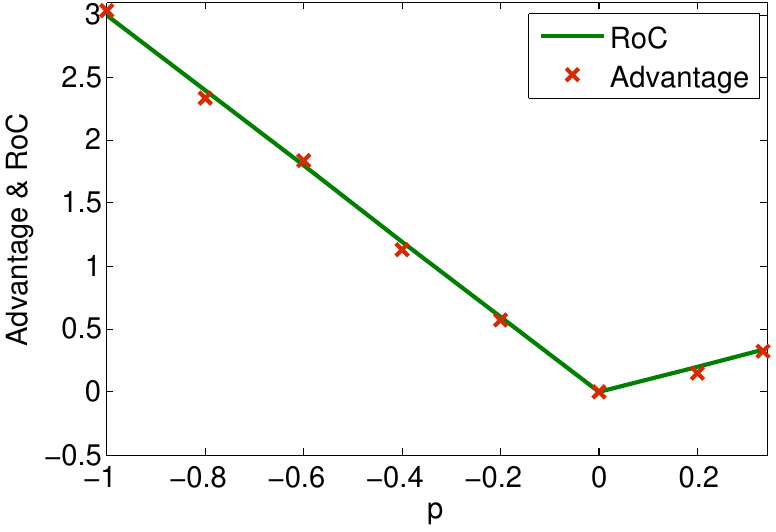}
	\caption{\label{fig:PDgame}
	\textbf{Experimental advantage of coherence for phase discrimination with an NMR system}~\cite{PengPRL2018}. The data are displayed in units of $\epsilon$, where $\epsilon\sim 10^{-5}$ denotes the polarization of the NMR system in the experimental condition. Reprinted with permission from W.~Zheng \emph{et al.}, Phys.\ Rev.\ Lett.\ \textbf{120}, 230504 (2018).%
		}
\end{figure}

The aforementioned approach was used in~\cite{PengPRL2018} to estimate the robustness of coherence with a coherence witness implemented in a multiqubit NMR system. Moreover, the experimental results of Ref.~\cite{PengPRL2018} explicitly verify an operational aspect of the robustness: its application in phase discrimination~\cite{PhysRevLett.116.150502}. In this task, a quantum system in the state $\rho$ enters into a ''black box'' implementing with probability $1/d$ a unitary operation $U_{\phi_k}=\exp(i N \phi_k)$,  where $N=\sum_{j=0}^{d-1} j\ketbra{j}{j}$ and $\phi_k = \frac{2\pi k}{d}$. The goal of phase discrimination is to guess which unitary $U_{\phi_k}$ was applied. For this purpose, one can implement a generalized measurement with measurement operators $\{M_k\}$ on the output state. The maximal success probability for correctly guessing the unitary $U_{\phi_k}$ in this procedure is given by 
\begin{equation}
    p_{\mathrm{succ}}(\rho)=\max_{M_k}\frac{1}{d}\sum_{k=0}^{d-1}\Tr[U_{\phi_k}\rho U_{\phi_k}^\dag M_k]. 
\end{equation}
As was shown in~\cite{PhysRevLett.116.150502,ProofRobustness}, states with nonzero coherence lead to a larger success probability, when compared to incoherent states. Moreover, it holds that~\cite{PhysRevLett.116.150502,ProofRobustness}
\begin{equation}
    \frac{p_\mathrm{succ}(\rho)}{p_\mathrm{succ}(\mathcal{I})} = 1 + C_R(\rho),
\end{equation}
where $p_\mathrm{succ}(\mathcal{I})$ is the success probability maximized over all incoherent input states. Thus, the robustness of coherence captures exactly the advantage achievable by having coherence in the input state~\cite{PhysRevLett.116.150502,ProofRobustness}.
This was explicitly verified in a two-qubit experiment carried out on an NMR quantum simulator in~\cite{PengPRL2018}, demonstrating the relevance of quantum coherence for experimental phase discrimination. The experimental results presented in~\cite{PengPRL2018} are shown in Fig.~\ref{fig:PDgame}.

\begin{figure}
	\centering
	\includegraphics[scale=1.2]{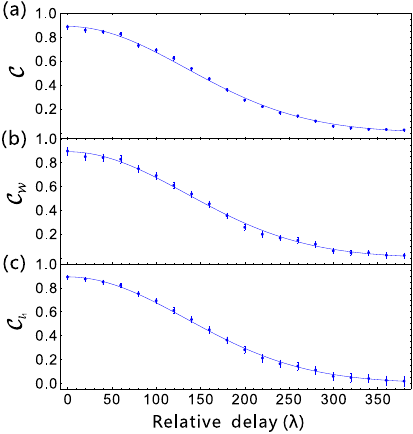}
	\caption{\label{fig:fringe1}
		\textbf{Experimental coherence estimation with linear optics}~\cite{wang2017directly}.  Results of the three coherence-measurement methods based on interference fringes, coherence witnesses, and state tomography are shown in (a), (b), and (c), respectively. $\lambda$ is the difference in optical path between $H$ and $V$ polarized photons in the quartz crystal, with a decohering coefficient $\kappa(\lambda)=\exp(-\frac{\Delta n^2l^2\sigma^2}{2c^2})$. Reprinted with permission from Y.-T.\ Wang \emph{et al.}, Phys.\ Rev.\ Lett.\ \textbf{118}, 020403 (2017).
	}
\end{figure}

Another experimental method for evaluating the robustness of coherence with linear optics was considered in Ref.~\cite{wang2017directly}. Notably, this work explicitly compares two methods for evaluating the robustness: one is an interference-fringe method based on the original definition of $C_R$ in Eq.~\eqref{eq:robustness_primal}, and the other is the witness approach discussed in Eq.~\eqref{eq:robustness_dual}. In the former method, the following two states are prepared in a linear optics setup:
\begin{align*}
    \rho=\frac{1}{2}[I+r_\rho(\sin\theta_\rho\cos\phi_\rho\sigma_x+\sin\theta_\rho\sin\phi_\rho\sigma_y+\cos\theta_\rho\sigma_z)], \\
    \tau=\frac{1}{2}[I+r_\tau(\sin\theta_\tau\cos\phi_\tau\sigma_x+\sin\theta_\tau\sin\phi_\tau\sigma_y+\cos\theta_\tau\sigma_z)].
\end{align*}
Robustness of coherence is then determined by experimentally finding the minimal value $s$ such that the interference fringes of the mixed state $(\rho+s\tau)/(1+s)$ vanish (cf.\ Eq.~\eqref{eq:robustness_primal}). In the latter method, robustness is estimated by measuring the expectation values $\Tr(\rho W)$ of suitable witness operators. Noting that $C_R(\rho) = C_{\ell_1}(\rho)$ for any single-qubit system~\cite{ProofRobustness}, one can then compare three experimental quantitative approaches to measuring coherence: the interference fringe method, the witness-based approach, and computation of the $\ell_1$-norm of coherence through full state tomography. The experimental implementation of this comparison, as performed in Ref.~\cite{wang2017directly}, is illustrated in Fig.~\ref{fig:fringe1}.

Most schemes for coherence detection are not device-independent, they assume that the final measurement is realized on a trusted device. To overcome this issue, the authors of~\cite{WitnessUntrustedPRL2019} (see also~\cite{Untrusted2}) developed a measurement-device independent coherence detection scheme, which was termed \emph{measurement-device-independent coherence witness} (MDICW). The scheme uses additional decoy states in a known state, and is robust against any bias on measurement devices. An experimental demonstration of the protocol with linear optics has also been presented~\cite{WitnessUntrustedPRL2019}. While the conventional coherence witness requires a trusted measurement device, this is not needed for the MDICW method. Moreover, it requires only one measurement setting, which provides an alternative way for detecting coherence and other quantum resources in a practical manner.

In Ref.~\cite{dai2020experimentally}, the authors present a fidelity-based method to derive experimentally accessible lower bounds for measures of coherence. Indeed, this can be understood as the measurement of a witness constructed as $W = |\phi_{\max}| \id - \ketbra{\phi}{\phi}$ where $\ket\phi$ is any chosen pure state, and $\phi_{\max}$ denotes the coefficient of $\ket{\phi}$ with the largest magnitude in the basis $\{\ket{i}\}$. The authors explicitly show how the method can deliver observable lower bounds for monotones such as the convex roof of the $\ell_1$-norm of coherence, the geometric coherence,
and the coherence of formation (see Section~\ref{sec:QuantifyingCoherence} for their definitions). 
The results are demonstrated through an application to experimental data from leading experimental multi-qubit entangled states~\cite{huang_2011,yao_2012,wang_2016-1}.

In Ref.~\cite{yuan2020direct}, the authors develop a collective measurement scheme (based on one-dimensional photonic quantum walks) for directly measuring different coherence quantifiers, including $\ell_1$-norm of coherence and relative entropy of coherence. For single-qubit states with Bloch vector $\mathbf{r}=(r_x,r_y,r_z)$, the $\ell_1$-norm of coherence and the relative entropy of coherence can be represented by
\begin{align}
     &{C}_{{\ell }_{1}}(\rho )=\sqrt{{r}_{x}^{2}+{r}_{y}^{2}},\\
     &{C}_{r}(\rho )=h\left(\frac{1+\left|{r}_{z}\right|}{2}\right)-h\left(\frac{1+r}{2}\right),
\end{align}
where $h(x)=-x\,{\mathrm{log}\,}_{2}x-(1-x)\,{\mathrm{log}\,}_{2}(1-x)$ is the binary entropy and $r={({r}_{x}^{2}+{r}_{y}^{2}+{r}_{z}^{2})}^{1/2}$ is the Bloch vector length.
Then it is found in Ref.~\cite{yuan2020direct}
that both $C_r$ and $C_{\ell_1}$ can be expressed in terms of outcome probabilities from a collective Bell measurement with elements 
\begin{align}
&M_{1}=\ket{\psi^{+}}\bra{\psi^{+}},\,M_{2}=\ket{\psi^{-}}\bra{\psi^{-}},\\
&M_{3}=\ket{\phi^{+}}\bra{\phi^{+}},\,{M}_{4}=\ket{\phi^{-}}\bra{\phi^{-}}
\end{align}
acting on $\rho\otimes\rho$ with the Bell states $\left|{\psi }^{\pm }\right\rangle =(\left|01\right\rangle \pm \left|10\right\rangle )/\sqrt{2}$ and $\left|{\phi }^{\pm }\right\rangle =(\left|00\right\rangle \pm \left|11\right\rangle )/\sqrt{2}$
. Denoting $P_i=\Tr[M_i\rho\otimes\rho]$, we have
\begin{equation}
    \begin{array}{ll}{r}_{x}^{2}+{r}_{y}^{2}=2({P}_{1}-{P}_{2}),\\ | {r}_{z}| =\sqrt{2\left({P}_{3}+{P}_{4}\right)-1},\\ r=\sqrt{1-4{P}_{2}}.\end{array}
\end{equation}
When compared to full quantum state tomography, the collective measurement scheme presented in~~\cite{yuan2020direct} can significantly increase the precision for estimating both $\ell_1$-norm of coherence and relative entropy of quantum coherence. 

Collective measurements based on photonic quantum walks are also useful for other aspects of quantum physics. In Ref~\cite{hou2018deterministic}, the authors present new experimental technologies for the realization of nonprojective collective measurement with a high fidelity over 0.99. In Ref.~\cite{tang2020experimental}, the authors apply collective measurements to optimal orienteering protocols based on parallel spins and antiparallel spins. Experimental results also show that such technologies can be used to reduce the backaction of projective measurements in quantum thermodynamics~\cite{wu2019experimentally,minimizingWu}.

Although the methods discussed above constitute the leading approaches for estimating the coherence of single quantum systems, a potential downside associated with their application is that implementing global measurements can be difficult when dealing with multi-qubit systems. For such systems, a spectrum-based estimation method was proposed in Ref.~\cite{yu_2019} which can be used to provide bounds for coherence measures such as the robustness of coherence $C_R$, the relative entropy of coherence $C_r$, or the $\ell_1$-norm of coherence $C_{\ell_1}$. The crucial aspect of this scheme is that a small number of local measurements can suffice to provide useful bounds. This was recently experimentally demonstrated in Ref.~\cite{ding_2020} with an implementation of the spectrum estimation scheme in a photonic setup. Therein, the authors generate highly entangled three- and four-qubit states --- the Greenberger-Horne-Zeilinger state, W state, and cluster state --- and use stabilizer measurements to ensure that measuring few local observables can effectively and efficiently bound multipartite coherence. In~\cite{Omran570,Song574} the authors generated GHZ states of up to 20 qubits using Rydberg atoms and a superconducting quantum processor, respectively. In the setup used in~\cite{Omran570}, the $\ell_1$-norm of coherence of the experimentally obtained state is related to the experimental GHZ state fidelity. {Besides quantum coherence concerning discrete variables, coherence of continuous variables has also been experimentally studied recently~\cite{zhang2021quantifying,kang2021experimental}.}

\subsection{Quantum coherence and uncertainty relations}
{The coherence of quantum state depends on the choice of reference basis. A natural question to ask is: if we use two or more incompatible reference bases, is there a trade-off or uncertainty relationship between coherence measures based on different reference bases? Several different theoretical approaches to this problem have been studied}~\cite{singh2016uncertainty,uncertaintyCheng1,uncertaintyDolatkhah,uncertaintyFan1,UncertaintyLuo1,uncertaintyMu,uncertaintyRastegin,uncertaintyYuan1}.

The entropic uncertainty relations and the coherence-based uncertainty relations have been experimentally tested via an all-optical platform~\cite{entropicPRA2020}. The unilateral coherence, defined in a bipartite system~\cite{ConvertingPRL,PhysRevA.95.052106}, is quantified as
\begin{equation}
    C^{AB}_M(\rho^{AB})=S(\rho^B_M)-S(\rho^{AB}),
\end{equation}
where $\rho^B_M=\sum_i\Tr_A(M^A_i\rho^{AB})$ with $M^A_i=\ketbra{i}{i}$ denote a complete set of orthogonal basis of subsystem $A$, and $S$ is the von Neumann entropy.

For a $d^2$-dimensional bipartite state $\rho^{AB}$, and a complete set of $d + 1$ mutually unbiased bases (MUB) applied on subsystem $A$, the entropic uncertainty relations and coherence-based uncertainty relations in MUBs can be derived as~\cite{PhysRevA.91.042133,entropicPRA2020}
\begin{align}
    & \sum_{m=1}^{d+1} S(M_m|B)\geq \log_2 d+d S(A|B),\\
    & \sum_{m=1}^{d+1} C^{AB}_M(\rho^{AB})\geq \log_2 d+S(A|B),
\end{align}
where $S(M_m|B)=S(\rho_{M}^{AB})-S(\rho^B)$, and $\rho_{M}^{AB}$ is the state $\rho^{AB}$ after a local projective measurement $M_m^A$, i.e., $\rho_{M}^{AB}=\sum_mM_m^A\rho^{AB}M_m^A$.

The authors of~\cite{entropicPRA2020} perform a complete set of MUB measurements on one of the subsystems consisting of three Pauli-operator measurements. Quantum state tomography is then applied to reconstruct the density matrices of the initial states and the postmeasured states. The experimental data is then used to obtain the corresponding measurement probability, magnitude of the uncertainty, and the lower bounds in the uncertainty inequalities~\cite{entropicPRA2020}. Related results have been presented in~\cite{tradeoffPRALv}, where the authors perform an optical experiment investigating trade-off relations between coherence measures in two noncommuting reference bases. It is shown that the divergence between the uncertainty quantified by coherence in two noncommuting bases and the lower bound would be larger with the increase of entanglement, contrary to the entropic uncertainty relations. Both of the approaches~\cite{entropicPRA2020,tradeoffPRALv} provide insight into the understanding of uncertainty relations and their connection to quantum coherence.

\subsection{Quantum coherence and path information}
{A particle can exhibit the properties of both waves or particles when going through an interferometer. The properties of particles are characterized by how much information one can have about the path of the particles through the device. The wave-like properties determine the visibility of interference pattern. There is a trade-off between the properties of particles and waves: the stronger one is, the weaker the other.} This was theoretically studied in Refs.~\cite{path3,originpath2}, leading to the inequality
\begin{equation}\label{eq:originpath}
    D^2+V^2\leq 1,
\end{equation}
where $D$ is a measure of path information and $V$ is the visibility of the interference pattern. In particular, $D=2p_{max}-1$ where $p_{max}$ is the maximum probability for correctly guessing the path of the particle with an ideal detector, and $V=(I_{max}-I_{min})/(I_{max}+I_{min})$, where $I_{max}$ ($I_{min}$) denotes maximum (minimum) of the interference pattern. The problem of interferometric duality in multibeam experiments has also been studied intensely~\cite{path3,path4,path5,path6,path7,path8}. In particular, quantitative measures of wave properties and particle properties for multibeam interferometers have been studied in Refs.~\cite{path3,path4,path5}. The complementarity in the double-slit experiment has been investigated in Refs.~\cite{path6,path7}.

The trade-off relation is generalized to duality in the $N$-path interferometer between coherence and path information in Ref.~\cite{theorypath}, namely
\begin{equation}
    \left(P_s-\frac{1}{N}\right)^2+X^2\leq \left(1-\frac{1}{N}\right)^2,
\end{equation}
where $P_s$ denotes probability of successfully identifying the detector states 
quantifying the available path information via optimized measurements, and $X=(1/N)C_{\ell_1}(\rho)$. Denoting $D=(P_s-1/N)/(1-1/N)$ and $C=X/(1-1/N)$ (V), we get exactly the form in Eq.~(\ref{eq:originpath}). The authors of~\cite{theorypath} also give an entropic version of the coherence-path information duality relation,
\begin{equation}
    C_r(\rho)+H(M:D)\leq H(\{p_i\}),
\end{equation}
where $C_r$ is given in Eq.~(\ref{eq:relent}), $H(M:D)=H(D)+H(M)-H(\{p_{ij}\})$. Moreover, $M,\,D$ denote the variables corresponding to the measurement and detector states, respectively. They take values $i\in\{1,2,....,N\}$ corresponding to $\rho_i$, with probability $p_i$. The corresponding information content is given by $H(D)=H(\{p_i\})=-\sum_{i=1}^{N}p_i\log p_i$.

Relation between coherence and path information is experimentally tested based on linear optical systems in~\cite{yuan2018experimental,gao2018experimental}, where photonic quantum states encoded in path or polarization are used. The wave property, quantified by the $\ell_1$-norm of coherence and the relative entropy of coherence, {can be experimentally obtained via tomography of the target state. The particle property, quantified by the path information, can be obtained via the discrimination of detector states, which is encoded in the polarization degree}. Both works~\cite{yuan2018experimental,gao2018experimental} suggest that coherence measures are promising candidates for the generalization of the interference visibility to describe the wave property in a multipath interferometer, deepening the understanding of coherence and providing new perspectives regarding wave-particle duality.

\subsection{Quantum coherence in distributed scenarios}\label{sec:distributed} 

Besides the studies on uncertainty relations of coherence in multipartite systems, there is a growing interest in quantum coherence as a resource in distributed scenarios~\cite{StreltsovPhysRevX.7.011024,ChitambarPhysRevLett.116.070402,ChitambarPhysRevLett.117.020402}, especially when it comes to state transformation and possible resource conversion. {Local operation and classical communication (LOCC) constitute one of the most important concepts in entanglement theory~\cite{HorodeckiRevModPhys.81.865}. They describe all the transformations that can be performed when two separated parties apply local quantum measurements and can only use a classical communication channel}. This concept has been extended to the resource theory of coherence, by introducing \emph{local quantum-incoherent operations and classical communication} (LQICC)~\cite{ChitambarPhysRevLett.116.070402,StreltsovPhysRevX.7.011024}. Relevant tasks which are based on LQICC operations include assisted coherence distillation~\cite{ChitambarPhysRevLett.116.070402,StreltsovPhysRevX.7.011024,Regula1807.04705,Oneshotassisted2}, where Alice helps Bob to extract quantum coherence locally.
The maximal rate of maximally coherent qubit states $\ket{+}$ on Bob's side is given by the distillable coherence of collaboration~\cite{ChitambarPhysRevLett.116.070402,StreltsovPhysRevX.7.011024}:
\begin{equation}\label{coc}
C_{d}^{A|B}(\rho)=\sup\left\{ R:\lim_{n\rightarrow\infty}\left(\inf_{\Lambda}\left\Vert \Lambda\left(\rho^{\otimes n}\right)-\ket{+}\!\bra{+}^{\otimes \lfloor Rn \rfloor}\right\Vert _{1}\right)=0\right\} ,
\end{equation}
where the infimum is taken over all $\textmd{LQICC}$ operations $\Lambda$. The distillable coherence of collaboration is bounded above as~\cite{ChitambarPhysRevLett.116.070402}
\begin{equation}\label{upperbound}
C^{A|B}_{d}(\rho^{AB}) \leq C^{A|B}_{r}(\rho^{AB}).
\end{equation}
Here, $C^{A|B}_{r}$ is the \emph{Quantum-Incoherent (QI) relative entropy}~\cite{ChitambarPhysRevLett.116.070402}, given as $C^{A|B}_{r}(\rho^{AB})=S(\Delta^B(\rho^{AB}))-S(\rho^{AB})$, and $\Delta^{B}$ denotes dephasing on Bob's side in the incoherent basis.

Experimental assisted coherence distillation in the single-copy regime has been performed in~\cite{wu2017experimentally}, based on a linear optical system. If Alice and Bob share a pure two-photon state, then Alice can help Bob to increase his local coherence in the single-copy setup, and the maximal relative entropy of coherence Bob can get is given by $S(\Delta(\rho^B))$, as has also been shown experimentally in~\cite{wu2017experimentally}.  The authors of~\cite{wu2017experimentally} also consider assisted coherence distillation of mixed states, showing that in some cases single-copy protocols can reach efficiencies close to the asymptotic upper bound in Eq.~(\ref{upperbound}).

Quantum coherence has also been investigated in the context of quantum nonlocality and steering~\cite{nonlocaltheory1}. For this, the authors of~\cite{nonlocaltheory1} derive a set of inequalities for various measures of quantum coherence. A violation of any of these inequalities implies that one party can steer quantum coherence of another remote party beyond what would be achievable with a single system and LOCC~\cite{nonlocaltheory1}. This phenomenon is called nonlocal advantage of quantum coherence~\cite{nonlocaltheory1}. The inequalities derived in~\cite{nonlocaltheory1} further allow to conclude that the underlying quantum state is steerable. Based on these theoretical results, the authors of~\cite{nonlocalPRA2019} report an experimental investigation of nonlocal advantage of quantum coherence for two-qubit states in an all-optical setup. Quantum state tomography is applied to reconstruct the initial and the measured states. The nonlocal advantage of quantum coherence is demonstrated for different coherence measures, including $\ell_1$-norm of coherence, relative entropy of coherence, and a coherence measure based on the skew information~\cite{nonlocalPRA2019}.

\begin{figure*}
	\centering
	\includegraphics[scale=0.35]{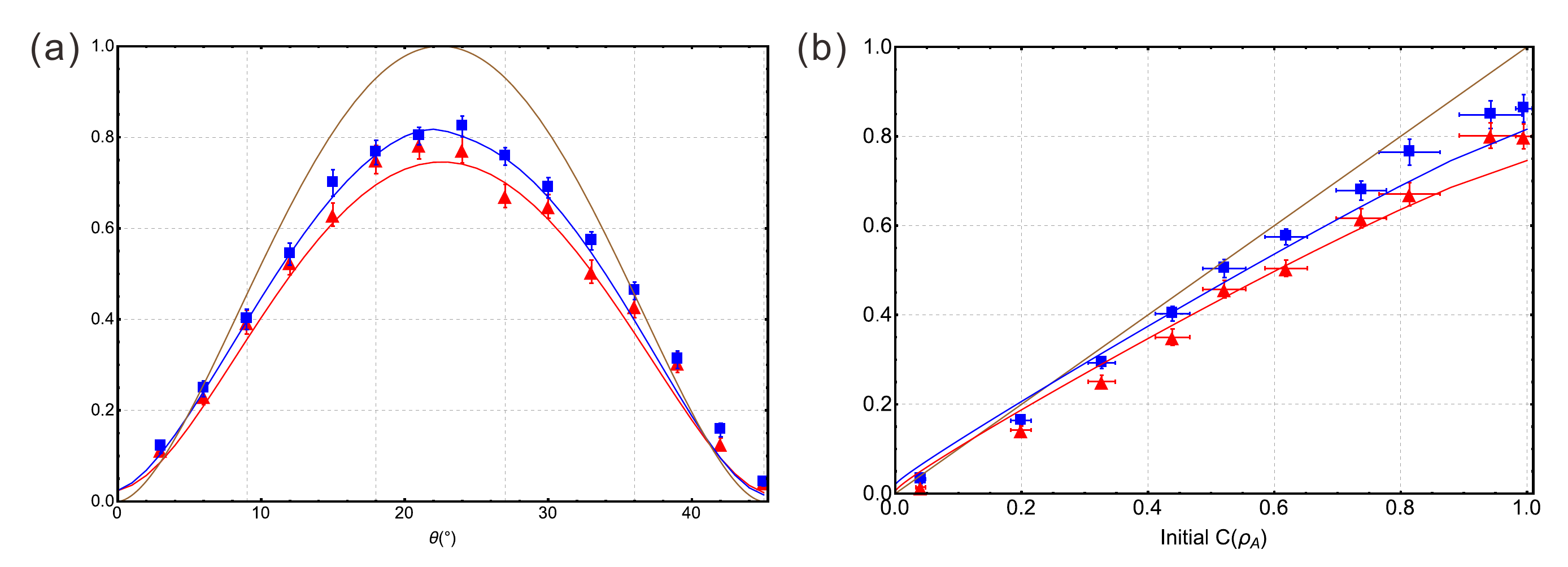}
	\caption{\label{fig:inter1}
		\textbf{Experimental inter-conversion of coherence and quantum correlations with a bulk optical system}~\cite{WuPRL2018}.  (a) The initial states on $A$ are prepared as $\ket{\phi}=\cos2\theta\ket{H}+\sin2\theta\ket{V}$. The experimentally obtained amount of quantum correlations (blue) and restored coherence (red) are plotted as functions of $\theta$. Solid lines represent fitting curves with experimentally reconstructed optical CNOT gate. (b) The initial states on $A$ are prepared as $\rho=\frac{1+a}{2}\ketbra{+}{+}+\frac{1-a}{2}\ketbra{-}{-}$. The experimentally obtained converted quantum correlations and restored coherence are plotted as functions of the initial coherence in $A$, which are experimentally obtained by state tomography. Reprinted with permission from K.-D. Wu \emph{et al.}, Phys.\ Rev.\ Lett.\ \textbf{121}, 050401 (2018).
	}
\end{figure*}

Distribution of quantum coherence in multipartite system has been studied in~\cite{Distribution2016,ConvertingPRL}. The authors of~\cite{Distribution2016} find theoretically that the metric properties allow quantum coherence to be decomposed into various contributions, which arise from local and intrinsic coherence. Several trade-off relations between the various contributions of coherence are found, and an experimental test of these relations has also been reported recently~\cite{ding2020experimental}.

The fact that both coherence and entanglement arise from the superposition principle of quantum mechanics suggests an equivalence between the resource theories of coherence and entanglement. A theoretical investigation of this equivalence shows that coherence can be converted into entanglement via incoherent operations, leading also to a correspondence between coherence and entanglement quantifiers~\cite{PhysRevLett.115.020403,ZhuhuangPRA2017}. This equivalence has also been explored for the resource theories of entanglement and superposition~\cite{Nathan,PhysRevLett.119.230401,qiao2018entanglement}, and for coherence and general quantum correlations beyond entanglement~\cite{ConvertingPRL}.

{In Ref.~\cite{WuPRL2018}, the authors illustrate a cyclic scheme where coherence initially in a quantum system $A$ is locally consumed to create an equal amount of quantum correlations with an incoherent ancillary system $B$.} The relation between consumed coherence and generated quantum correlations reads~\cite{WuPRL2018}
\begin{equation}\label{discordbound}
D(\rho^{AB}_{I})=D[\Lambda^{AB}(\rho^A\otimes\tau^B)]\leq C_r(\rho^{A}).
\end{equation}
Here, $\rho^A$ is the initial quantum state of $A$, $\tau^B$ is an incoherent state of $B$, and $\Lambda^{AB}$ represents a bipartite incoherent operation. Moreover, $D(\rho^{AB})=\min_{\sigma^{AB}\in\mathcal{CC}}S(\rho^{AB}\|\sigma^{AB})$ is a quantifier of quantum correlations beyond entanglement~\cite{ModiPhysRevLett.104.080501,RMPdiscord} with $\mathcal{CC}$ denoting the set of classically correlated states, i.e., states of the form 
\begin{equation}
    \sigma^{AB}=\sum_{ij}p_{ij}\ketbra{e^A_i}{e^A_i}\otimes\ketbra{e^B_j}{e^B_j},
\end{equation}
where $\{\ket{e^A_i}\},\,\{\ket{e^B_i}\}$ are orthogonal bases for $A$ and $B$. The upper bound in Eq.~(\ref{discordbound}) can be saturated and we denote with $\rho^{AB}_{\mathrm{Imax}}$ the state with maximum quantum correlations. These correlations are then harnessed to restore coherence in $A$, i.e.,
\begin{equation}\label{cohbound}
C_r(\rho^{A}_{\textmd{LQICC}})\leq D(\rho^{AB}_{\mathrm{Imax}}) = C_r(\rho^{A}),
\end{equation}
where $\rho^{A}_{\textmd{LQICC}}=\sum_jp_j\rho^A_j$, $p_j=\Tr \left(\bra{e^B_j}\rho^{AB}_{\mathrm{Imax}}\ket{e^B_j}\right)$ and $\rho^A_j=\bra{e^B_j}\rho^{AB}_{\mathrm{Imax}}\ket{e^B_j}/p_j$.

Under ideal conditions, the cycle described above is lossless, and can be repeated infinitely many times. The authors of~\cite{WuPRL2018} report one round of this cycle based on a bulk optical system, showing explicitly how coherence encoded within a photonic qubit can be converted into quantum correlations between it and an ancilla via incoherent operations. By measurement of the ancillary photon, up to 80 percent of the coherence within the original qubit is restored. The results presenting converted quantum correlations and restored coherence are shown in Fig.~\ref{fig:inter1}. Other experimental tests of related resource conversion protocols have been investigated in~\cite{wang2019witnessing,interPRA}.

A protocol for witnessing spatial quantum superposition between two distant parties using only local measurements has been proposed in~\cite{CoherenceEquality}, opening the possibility of detecting coherence without interference patterns.

\subsection{Single-shot state conversion and coherence distillation}

Our discussion in Section~\ref{sec:distributed} focused on the asymptotic rates of coherence transformations. Although such quantities are useful as the ultimate bounds on achievable transformations, the idealized assumption of coherently manipulating an unbounded number of quantum systems can mean that these rates are far from what is practically realizable in a lab. It is then of interest to understand precisely the performance of coherence manipulation at the level of practical single- or multi-shot protocols. Broadly speaking, the analysis of non-asymptotic coherence manipulation explicitly takes into consideration the fact that many desired state transformations can be impossible to perform in practice under the action of a chosen class of free operations. One can then consider \textit{probabilistic} protocols, in which the transformation $\rho \to \rho'$ is achieved only with some probability $p$ but fails with probability $1-p$, or \textit{approximate} protocols, in which the output state of the protocol, $\tau$, is not exactly the desired target state $\rho'$ but only a state close to it, and the transformation error can be measured e.g.\ by the infidelity $1-F(\tau, \rho')$~\cite{Imaginarity1,Imaginarity2}.

An important example of such a task is one-shot coherence distillation, in which the target state is the maximally coherent state $\ketbra{+}{+}$ or several copies thereof. On the theoretical side, the single-shot investigation of this problem has been approached using both approximate~\cite{Regula1711.10512,zhao_2019,lami_2019-1} and probabilistic~\cite{PhysRevLett.121.070404} transformations. This task can be investigated also in the distributed scenario described above, where two parties are collaborating to distill maximal coherence on one side~\cite{Vijayan1804.06554,Regula1807.04705}.

The first experimental realization of approximate (deterministic) coherence distillation was performed in an optical setup in Ref.~\cite{xiong_2019}. There, the authors introduce a method of transforming three- and four-dimensional pure quantum states into states which approximate a two- or three-dimensional maximally coherent state. The higher-dimensional transformations are realized by a multi-step protocol divided into a sequence of SIO operations, each acting on two qubits at a time. 
Furthermore, the assisted coherence distillation experiment performed in Ref.~\cite{wu2017experimentally} is also an example of a single-shot realization of an assisted distillation protocol, although the quantitative measures employed there are more closely related to the asymptotic rates.

More generally, one can study transformations of quantum states beyond the task of distillation. Ref.~\cite{wu2020quantum} characterized general probabilistic one-shot transformations of single-qubit systems under (strictly) incoherent operations and showcased the optimal protocols in an explicit experimental implementation. The insights were also extended to the problem of state conversion in assisted (distributed) scenarios, obtaining exact optimal solutions for relevant cases of states.

\subsection{Dynamics of quantum coherence}
Coherence is a vulnerable quantum resource, unavoidably vanishing at macroscopic scales of space, time, and temperature~\cite{ZurekRevModPhys.75.715,SchlosshauerRevModPhys.76.1267}. {This becomes particularly clear when one studies the dynamical behavior of coherence in the presence of dissipation, where the system is rarely isolated and usually loses its information to its environment.}

{Numerous studies have investigated the dynamics of quantified coherence in open quantum systems~\cite{chanda2016delineating,he2017non,mirafzali2019non,zhang2015role,coherencetrappingPRA,passos2019non,man2018temperature,ccakmak2017non,yu2018quantum,bhattacharya2016effect,PhysRevA.99.022107}. The authors of~\cite{chanda2016delineating} proposed a class of protocols for detecting and quantifying the non-Markovianity of incoherent open system dynamics (IOSD). This approach is broadly based on the fact that any valid coherence measure monotonically decreases under incoherent operations (recall Sec.~\ref{sec:QuantifyingCoherence}), which means that coherence can be used to detect the non-Markovianity for quantum dynamics which preserve the set of all incoherent states, i.e., IOSD. Consider a quantum system described by a state $\rho$ undergoing an IOSD characterized by $\Lambda_{t}$. Let $\rho_{s}$, $\rho_t$ denote the quantum state after evolution time $s$ and $t$, respectively. For any coherence measure $C$, the amount of quantum coherence decreases monotonically:
\begin{equation}
C\left(\rho_t\right)\leq C\left(\rho_s\right), \forall t>s>0.
\end{equation}
A similar approach has also been used with coherence measures based on the Tsallis relative $\alpha$ entropies~\cite{mirafzali2019non}. }

In~\cite{he2017non} the authors proposed an alternative non-Markovianity measure based on the relative entropy of coherence in a bipartite setup. Consider a bipartite system $AB$ in a state $\rho^{AB}$, where only the system of Alice undergoes an IOSD characterized by $\Lambda_{t}$. Let $\rho^{AB}_s$ and $\rho_t^{AB}$ denote the overall quantum state after evolution time $s$ and $t$, respectively. The relative entropy of coherence of the total system decreases monotonically, i.e.,
\begin{equation}
C_r\left(\rho^{AB}_t\right)\leq C_r\left(\rho^{AB}_s\right), \forall t>s>0,
\end{equation}
{from which the proposed measure can effectively capture the features of non-Markovianity of open quantum systems.}

While the methods just described allow to detect non-Markovianity only for incoherent evolutions, the tools presented in~\cite{wu2020detecting} allow to characterize non-Markovianity of any evolution based on the $\mathcal{QI}$ relative entropy (see discussion below Eq.~\eqref{upperbound}). The authors of~\cite{wu2020detecting} also experimentally investigated the evolution of the local coherence, the extended coherence, the $\mathcal{QI}$ relative entropy {of both the system of interest and the ancillary particle}, and the steering induced coherence {of the ancillary particle} in both Markovian and non-Markovian processes. In the optical experiments, the polarization photonic state is regarded as the open system and the frequency of the corresponding photons is used as the environment. The open system dynamics is then realized by the coupling between the polarization and a birefringence crystal, where the non-Markovianity depends on the spectral properties of the photons. The experimental results in~\cite{wu2020detecting} show that information back-flow can be experimentally detected with respect to different reference bases.

The authors of~\cite{environmentYu} investigated the sudden change phenomena of quantum coherence both theoretically and experimentally. In particular, the authors use decoherence in the polarization basis of a single photon, where the open system is determined by the {difference in evolved phases} at every spatial position. Coherence dynamics occurs when the system and environmental (spatial modes) interact in {several tilted} birefringent media. Thus by tuning the graphics of the environment, the sudden death phenomenon of quantum coherence can be observed~\cite{environmentYu}. Related experimental results investigating the dynamical behavior of quantum coherence in open quantum systems have been reported in~\cite{passos2019non,PhysRevA.99.022107,HuanFragility}.

\begin{figure*}
	\centering
	\includegraphics[scale=0.38]{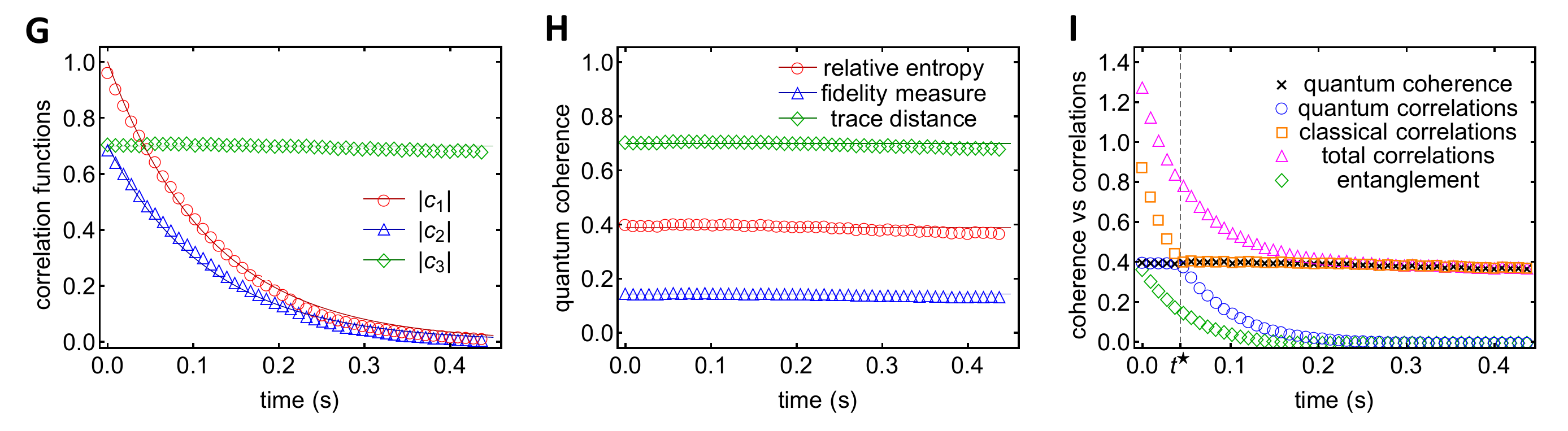}
	\caption{\label{fig:frozen1}
		\textbf{Experimental freezing of quantum coherence with an NMR setup}~\cite{coherencefreeze2016}. (G) Experimental values of $|c_j|$ (points), where $c_j=\langle \sigma_j\otimes\sigma_j\rangle$, along with theoretical predictions (solid lines) based on phase damping noise. (H) Experimental observation of time-invariant coherence, measured by relative entropy of coherence (red circles), geometric coherence (blue triangles), and a coherence measure based on the trace distance (green diamonds). (I) Experimental dynamics of coherence and different correlation measures. Figure is taken from~\cite{coherencefreeze2016}, sub-figures (A-F) are omitted. Reprinted with permission from I. A. Silva \emph{et al}., Phys.\ Rev.\ Lett.\ \textbf{117}, 160402 (2016). 
	}
\end{figure*}

Apart from the relations between coherence and non-Markovianity, one of the most interesting phenomena observed in the context of coherence dynamics is the possibility for its freezing, which was first put forward in~\cite{theoryFrozen1} and developed further in~\cite{theoryFrozen2,FrozenCoherenceAtom}. The authors of~\cite{theoryFrozen1} demonstrate that the $\ell_1$-norm of coherence is frozen for some initial states under certain incoherent evolutions. Interestingly, in contrast to the single-qubit case, universal freezing of quantum coherence occurs in multi-qubit systems with an even number of qubits. When initialized in a certain class of states, such systems demonstrate measure-independent freezing of coherence under a certain class of incoherent qubit dynamics~\cite{theoryFrozen1}.

The phenomenon of frozen quantum coherence has been demonstrated experimentally in a room temperature NMR setup~\cite{coherencefreeze2016}. In particular, after preparation in a generalized Bell diagonal state, the multiqubit ensemble is left to evolve under phase damping noise. Constant coherence is then observed within the experimentally considered time scales up to $1$ second. Experimental freezing of coherence is demonstrated explicitly for the relative entropy of coherence, the geometric coherence, and a coherence measure based on the trace distance. Moreover, the authors of~\cite{coherencefreeze2016} prove theoretically that coherence can decay in the case of more general initial states, while it remains above a guaranteed threshold at any time. This phenomenon is also experimentally observed with two-qubit states. In Fig.~\ref{fig:frozen1}, which is taken from~\cite{coherencefreeze2016}, we show the experimentally obtained evolution of correlation functions, coherence, and correlation quantifiers. 

Experimental freezing of coherence has also been investigated in~\cite{SelfguideYuPRA}. The authors of~\cite{SelfguideYuPRA} develop a self-guided quantum coherence freezing method, which can guide either the quantum channels or the initial state to the coherence-freezing zone from any starting estimate, and investigate the protocol using linear optics.

\alex{Another line of research explores practical applications of frozen
or resilient coherence in quantum parameter estimation~\cite{giovannetti2004quantum,MetrologyPRL1,giovannetti2011advances}. The authors of~\cite{ZhangPRL2019} use a highly controllable photonic system to study the resilience of quantum coherence and metrology against transversal noise. In particular, they demonstrate frozen quantum coherence in a 4-photon GHZ state prepared in both the computational and $\sigma_x$ bases and then subjected to Markovian bit-flip noise. The phenomenon that quantum Fisher information
for estimating a phase encoded along the $\sigma_z$ basis is also
frozen in the GHZ state prepared in the $\sigma_x$ basis is observed. Moreover, a frequency estimation task with additional bit-flip noise is considered, and it is shown that one can surpass the standard quantum limit using GHZ probes of up to 6 qubits, despite their exposure to noise of comparable strength to the signal~\cite{ZhangPRL2019}. Results in~\cite{relatedPRL1} demonstrate that local encoding provides an advantage for phase estimation in the presence of noise.}



\subsection{Multilevel coherence}

Most of the experimental works on coherence quantification and applications concerns coherence in qubit systems. Since higher-dimensional coherence has a much richer structure, a precise understanding of its properties requires one to be able to distinguish between the basic two-level coherence commonly found in qubit systems and the fine-grained structure of superposition in higher-dimensional systems. The quantification and certification of such multilevel forms of quantum coherence has been studied in Refs.~\cite{witt_2013,levi_2014,chin_2017,chin_2017-1,regula_2018-2,PRX2018certification}. In particular, the multilevel nature of coherence can be characterized by the number $r_C(\ket{\psi})$ of nonzero coefficients $c_i$ for a pure state $\ket{\psi}=\sum_i c_i\ket{i}$ written in a given reference basis $\{\ket{i}\}$, which can be suitably extended to mixed states by defining $\mathcal{C}_{k}$ as the set of convex combinations of states with $r_C(\ket\psi) \leq k$. Specifically, $\mathcal{C}_k$ is then the set of $(k+1)$-level coherence-free states, with $\mathcal{C}_1$ reducing to the standard notion of two-level coherence that we discussed before. 

Within such a framework, analogous definitions of multilevel coherence-free operations and $k$-decohering operations can be introduced~\cite{levi_2014,PRX2018certification}. Mirroring the two-level cases, a number of $(k+1)$-coherence monotones can also be defined~\cite{chin_2017-1,regula_2018-2,PRX2018certification,regula_2018}. Of particular importance is the robustness of ($k+1$)-level coherence $R_{C_k}$, defined as the minimal amount of noise that has to be added to a state to destroy all ($k+1$)-level coherence~\cite{PRX2018certification}:
\begin{equation}
    R_{C_{k+1}}(\rho):=\inf_{\tau\in\mathcal{D}(\mathcal{H})}\left\{s\geq 0\,:\,\frac{\rho+s\tau}{1+s}\in \mathcal{C}_k\right\}.
\end{equation}
This measure satisfies all properties desired from a valid resource monotone also in the multilevel setting~\cite{PRX2018certification}. It is computable with a semidefinite program in general, and an explicit expression can be obtained for all pure states $\ket\psi$~\cite{regula_2018}. However, when the full tomographic knowledge of a given state cannot be assumed, many of such methods become experimentally infeasible.

\begin{figure}
	\centering
	\includegraphics[scale=0.05]{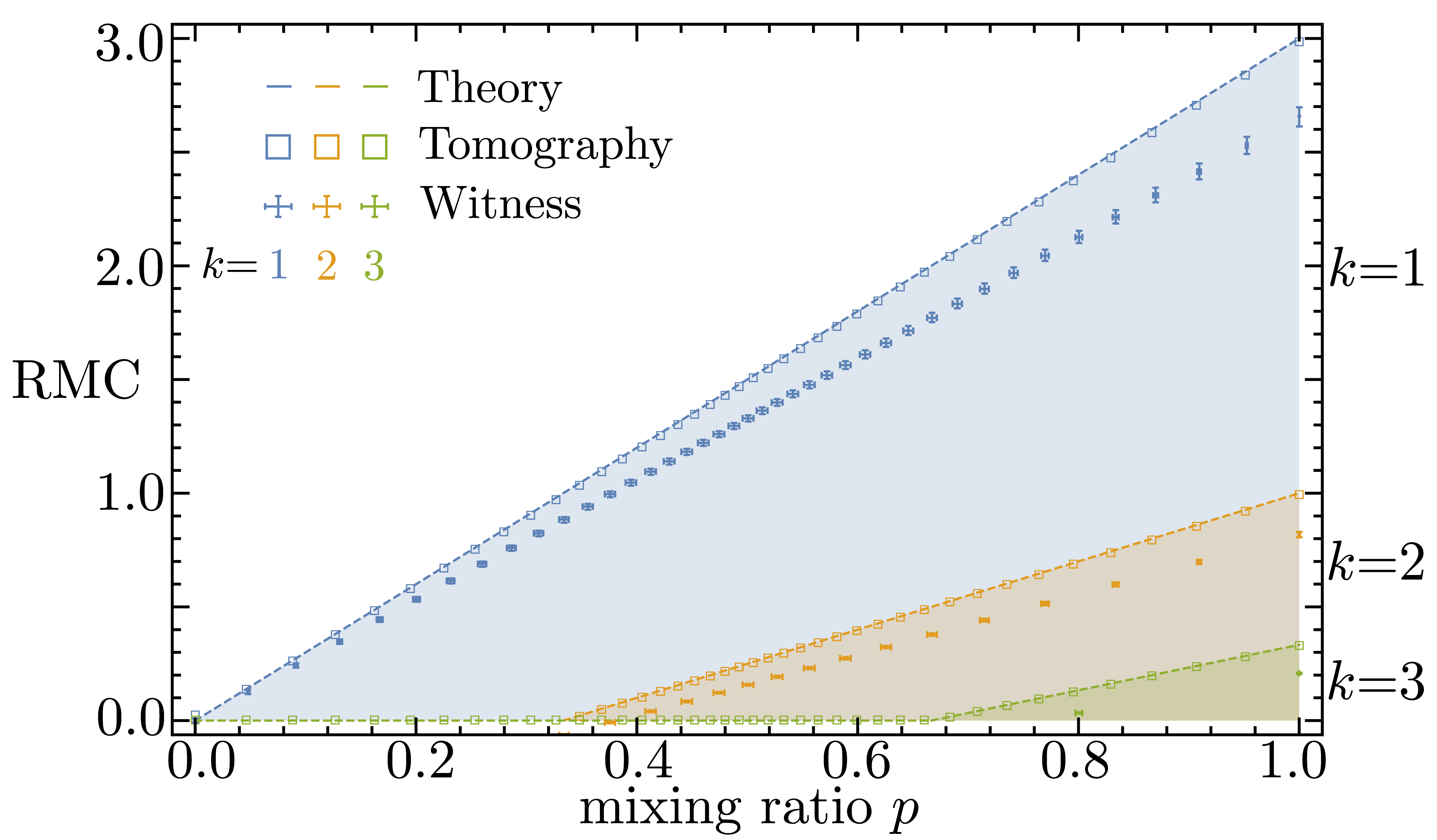}
	\caption{\label{fig:Multilevel}
		\textbf{Experimental quantification of multilevel coherence}~\cite{PRX2018certification}. The (k+1)-level coherence for the class of states in Eq.~(\ref{eq:noisycoherentstate}) is shown with $p\in[0,1]$, as measured by the robustness of multilevel coherence (RMC). The open squares denote the robustness of (k+1)-level coherence estimated from  the experimentally reconstructed density matrices, and the data points with error bars show the absolute values of the negative expectation values of the coherence witness. Reprinted with permission from M. Ringbauer \emph{et al}., Phys. Rev. X \textbf{8}, 041007 (2018).
	}
\end{figure}

Ref.~\cite{PRX2018certification} studied the experimental quantification and certification of coherence in this setting. The authors illustrated the theoretical framework in practice using a single-photon interferometer with varying
degrees and levels of coherence. Specifically, a family of noisy maximally coherent states was considered, 
\begin{equation}\label{eq:noisycoherentstate}
    \rho(p)=(1-p)\frac{I}{d}+p\ketbra{\psi^+_d}{\psi^+_d},
\end{equation}
where $d$ denotes the dimension ($d=4$ in the experiment) and $\ket{\psi^+_d}$ is a $d$-dimensional maximally coherent state. Such states form a structurally simple testbed for the investigation of coherence in the multilevel setting, since varying the noise parameter $p$ allows for a smooth interpolation from maximal ($d$-level) coherence through all lower coherence levels. The authors demonstrate that the robustness $R_{C_{k+1}}(\rho)$ can be computed for such states by a suitable choice of $(k+1)$-coherence witnesses, extending the two-level case and enabling an efficient quantification of multilevel coherence as well as certification of a given coherence level (see Fig.~\ref{fig:Multilevel}).


\subsection{Quantum coherence of operations and measurements}

Various generalizations of the methods used to study the superposition of quantum states were proposed in order to understand the coherence of the dynamical aspects of state manipulation: quantum channels and measurements. In the context of the resource theory of quantum coherence, this was first studied in the form of coherence generating power~\cite{bendana_2017,bu_2017-3}, that is, the largest amount of coherence that can be created by a given operation, or by asking the reverse question of how much coherence is required to simulate the action of a given operation using only incoherent operations~\cite{bendana_2017,diaz_2018-2}. Following earlier studies of nonclassical properties of quantum channels~\cite{meznaric_2013}, a more general line of research was later proposed with the aim of understanding the intrinsic coherence of quantum operations~\cite{OperationTheory1,DynamicalCoherence2}. Frameworks suited to the coherence of quantum measurements have also been introduced~\cite{cimini_2019,baek_2020}.

Experimental demonstrations of these methods have been realized in several works. Ref.~\cite{cimini_2019} first considered a method of detecting the coherence of a quantum measurement based on computing the variance of the measurement statistics and comparing it with classically (incoherently) achievable data. The practicality of this approach was demonstrated using data from a single-qubit photonic experimental setup.
In Ref.~\cite{DetectorPRL2020}, the authors experimentally investigate a quantitative approach to characterizing the coherence of quantum operations proposed earlier in Ref.~\cite{OperationTheory1}. An improved algorithm for quantum detector tomography is derived, and is applied to reconstruct the positive-operator-valued measures of
a weak-field homodyne detector in different configurations, which are then employed to evaluate how well the setup can detect coherence using two computable measures. Finally, Ref.~\cite{baek_2020} proposed several other norm-based quantifiers of coherence of measurements, and investigated their practical computability through an evaluation on the IBM Q quantum processor.

\section{Conclusions} \label{sec:Conclusions}

The study of quantum coherence as a resource has found a firm foundation in the setting of quantum resource theories, leading to a rich and active investigation of ways to measure and transform coherence. This practically-motivated framework is relevant in any physical setting in which an experimenter is limited to quantum operations which cannot create or detect superposition. Various experimental platforms, such as linear optics and NMR systems, can be used to explicitly study the manipulation and characterization of quantum coherence in these settings. In this article, we reviewed recent experimental progress on the implementation of such protocols. We saw that a number of efficient approaches have been devised to quantify and detect coherence without requiring costly methods such as full-state tomography. We discussed schemes realizing coherence manipulation in distributed scenarios, coherence distillation, and the dynamical connections between quantum coherence and open quantum systems. We also saw that close connections can be established between coherence and other phenomena such as non-Markovianity, quantum discord, and quantum uncertainty relations, also explicitly studied in experimental settings. Though most of the recent works realize proof-of-principle demonstrations of frameworks in the resource theory of quantum coherence, these findings lay the groundwork for practical applications, and can specifically impact noisy quantum information processing and nanoscale technologies.

So far, the experimental progress has been limited to low-dimensional quantum systems, mostly single qubits. The promising developments in the technology to prepare higher-dimensional entangled states~\cite{bavaresco2018measurements,erhard2018experimental,high1,high2,high4} and multipartite entangled states with up to 20 qubits~\cite{Omran570,Song574} suggest that the experimental investigation of coherence, and in particular multi-level coherence in multipartite scenarios, might also expand in the very near future. However, the difficulty of maintaining and coherently manipulating superpositions of a large number of quantum systems remains a major practical difficulty, and overcoming it will be the main obstacle to efficient practical utilization of quantum coherence as a resource.

\begin{acknowledgements}
The work at the University of Science and Technology of China is supported by the National Key Research and Development Program of China (Nos. 2017YFA0304100 and 2018YFA0306400), the National Natural Science
Foundation of China (Grants No. 61905234, 11974335, 11574291, and 11774334), the Key Research Program of Frontier Sciences, CAS (Grant No. QYZDYSSW-SLH003), and the Fundamental Research Funds for the Central Universities (Grant No. WK2470000026). A. S. acknowledges financial support by the "Quantum Coherence and Entanglement for Quantum Technology" project, carried out within the First Team programme of the Foundation for Polish Science co-financed by the European Union
under the European Regional Development Fund.
\end{acknowledgements}


%

\end{document}